\documentclass[12pt]{article}
\usepackage{a4}
\usepackage{graphicx,amssymb,amstext,amsmath,pgfplots,tikz}
\usetikzlibrary{shapes,arrows}
\usepackage{tocloft,calc}
\begin{document}
\title{\textbf{ Developing and building an absorption imaging system for Ultracold Atoms}}
\author{\begin{small}Tadas Pyragius \end{small}\\ \textit{\begin{small}University of Nottingham, Department of Physics and Astronomy, NG7 2RD \end{small}} \\ \textit{\begin{small}ppytp@nottingham.ac.uk\end{small}}}
\maketitle
\begin{abstract}
In this report I attempt to outline the process of developing and building an absorption imaging system capable of imaging ultracold atoms. In the theory section I will discuss the elements required to estimate the atom number and derive the equations required to determine the temperature of the atom cloud. I will outline the general features of a microscopic -- two lens imaging setup via the principles of geometric optics. A discussion of optical design process and its features will be carried out with an example on absorption imaging. The essence of basic optical design process will be given.
\end{abstract}
\tableofcontents
\section{Introduction}
An optical imaging system is at the heart of almost every physics experiment. Efficient collection of light by an optical system and its transformation into an image can provide a great insight into various physical processes that take place. \\ \\ Discovery and advancement in laser cooling techniques has led to a rapid development of ultra -- low temperature physics where new states of matter were observed for the first time since it's prediction by Bose and Einstein in the early 20th century. \\ \\ One of the most striking predictions of quantum mechanics and Bose -- Einstein statistics is the phenomenon of Bose -- Einstein condensation, where a dilute and weakly interacting gas confined in an external potential would display quantum effects on a macroscopic scale. The need to observe such phenomena and test these theories has led to the development of various optical imaging techniques ranging from absorption to fluorescence and more recently light sheet detection. In this report however, we will only discuss absorption imaging. \\ \\ Absorption imaging is by far the most popular imaging technique used in ultracold atom experiments. It can provide information about the spatial distribution of atoms, the atom number and its average temperature. [1]
\section{Theory}
\subsection{Absorption imaging and estimation of atom number}
In absorption imaging a cold atom cloud is illuminated by a laser beam which is resonant with it. The absorption of light by the atoms casts a shadow which is then imaged onto a CCD camera. The amount of light absorbed by the cloud is the optical density and the corresponding effect on the intensity of the laser beam is given by Beer’s Law: \begin{eqnarray} I=I_{0}e^{-OD} \end{eqnarray} Where, $I_{0}$, is the intensity of the incident laser beam and, $I$, is the intensity of outgoing laser beam. If the incoming intensity is below saturation, then we can rewrite equation (1) in terms of the atom column density, $n(x,y)$, and the absorption cross section, $\sigma$ [1]: \begin{eqnarray} I=I_{0}e^{-n(x,y)\sigma} \end{eqnarray} Here the absorption cross section is given by: \begin{eqnarray} \sigma = \frac{\sigma_{0}}{1+4(\frac{\Delta}{\Gamma})^{2}+\frac{I_{0}}{I_{sat}}} \end{eqnarray} Where, $\sigma_{0}$, is the on resonance cross section and is given by [2]: \begin{eqnarray}\sigma_{0}=\frac{\hbar\omega\Gamma}{2I_{sat}} \end{eqnarray} Here, $\Delta$ is the detuning of the laser field from the atomic resonance, $\Gamma$, is the natural decay rate of the excited state and, $I_{sat}$, is the saturation intensity. \\ \\ Now we can equate equations (1) and (2): \begin{eqnarray} OD(x,y) = n(x,y)\sigma \end{eqnarray} Experimentally, we can measure the optical density by measuring the incoming and the outgoing intensities of the laser beams. However, to measure the actual optical density, we need to take into account the effects of saturation and dark currents as well as stray light. Taking all this into account, the final optical density is calculated using [3]: \begin{eqnarray}OD_{actual}=OD_{mod}+(1-e^{OD_{mod}})\frac{I_0}{I_{s}} \end{eqnarray} With $OD_{mod}$ given by: \begin{eqnarray}OD_{mod}=ln\frac{1-e^{-OD_{sat}}}{e^{-OD_{meas}}-e^{-OD_{sat}}} \end{eqnarray} And finally, the $OD_{meas}$ is given by:\begin{eqnarray}OD_{meas}=ln\frac{I_{light}-I_{dark}}{I_{atoms}-I_{dark}} \end{eqnarray} Where, $I_{light}$, is the intensity of light without the atoms, $I_{dark}$, is the intensity of the probe beam off and camera fully open and, $I_{atoms}$, is the intensity of light with atoms. \\ \\ Now, to estimate the atom number we rearrange the equation (5) and integrate along the $x$ and $y$ directions to get the total atom number: \begin{eqnarray}n = \int_{-\infty}^{\infty} \int_{-\infty}^{\infty} n(x,y) dxdy=\frac{1}{\sigma}\int_{-\infty}^{\infty} \int_{-\infty}^{\infty} OD_{actual}dxdy \end{eqnarray} Here the integration of the optical density (OD) implies that it's intensity components, $I$, are integrated across the image since the intensity varies spatially.
\subsection{Derivation of the temperature of the cloud}
We assume a non--interacting cloud that obeys Boltzmann's statistics. Such particles thus follow the Boltzmann distribution which states that the probability of a particle being in the $E_{i}^{th}$ energy state at temperature, $T$ is given by: \begin{eqnarray}p_{i}=\frac{1}{Z}exp{(-\frac{E_{i}}{k_{B}T})}dE \end{eqnarray} Where, $Z$, is the partition function and, $k_B$, is the Boltzmann constant. \\ \\ Now suppose we have $N$ such particles in the cloud, then, the mean number of particles in the $E_{i}^{th}$ energy state at temperature, $T$ is given by: \begin{eqnarray} n_{i}=\frac{N}{Z}exp{(-\frac{E_{i}}{k_{B}T})}dE \end{eqnarray} We assume that these particles have motions that can fit the classical picture. Thus, we can express the kinetic energy of such particle as: \begin{eqnarray}E_k=\frac{1}{2}mv^2 \end{eqnarray} We would now like to express our number distribution in terms of velocity differential instead of energy, this can be achieved by the following trick: \begin{eqnarray} E_k=\frac{1}{2}mv^2 \Rightarrow \frac{dE}{dv}=mv \Rightarrow dE=mvdv \end{eqnarray} Thus, by replacing $E$ and $dE$ in terms of velocity components and differentials we obtain: \begin{eqnarray} n_{i} = \frac{NmV}{Z}exp{(-\frac{mv^2}{2k_BT})}dv \end{eqnarray} In the time of flight experiments, (TOF), we cannot directly measure the velocities of each individuals atoms in the cloud. However, we can measure the overall expansion of the cloud in terms of its size in the $x$ and $y$ directions. If we assume that the cloud has an initial size of zero (treating it as a mathematical point), and no initial velocity, we can apply simple kinematics equation that relates the speed of the particle with the distance moved: \begin{eqnarray}v=\frac{x}{t} \end{eqnarray} Now we can replace the velocity, $v$, and its corresponding differential in equation (14) with equation (15) by applying identical method used to obtain equation (13), this gives the new distribution: \begin{eqnarray}n_{i}=\frac{Nmx^2}{Zt}exp{(-\frac{mx^2}{2k_BTt^2})}dt \end{eqnarray} The equation (16) has the form of a Gaussian distribution, namely: \begin{eqnarray}f(x)=\frac{1}{\sigma\sqrt{2\pi}}exp{(-\frac{x^2}{2\sigma^2})}dx \end{eqnarray} In the experiment, we can directly measure $x$ which is the variable. We can now equate the arguments in the exponentials of the equations (16) and (17): \begin{eqnarray}\frac{mx^2}{2k_BTt^2}=\frac{x^2}{2\sigma^2} \end{eqnarray} Cancelling the $x$ and rearranging for $\sigma$, we obtain the size of the cloud at a given time $t$ and temperature $T$: \begin{eqnarray} \sigma=\sqrt{\frac{k_BT}{m}}t \end{eqnarray} So in the TOF experiment, we can take a picture of the cloud at different times, $t$, and measure the corresponding cloud size. Then by estimating the gradient of the cloud size with respect to time we obtain the gradient which is the argument under the square root. This can then be rearranged algebraically to obtain the temperature of the cloud. If the cloud has initial size, $\sigma_0$, then the time evolution of the cloud size is just the sum of the squares: \begin{eqnarray} \sigma(t) = \sqrt{\frac{k_BT}{m}t^2 + \sigma_0^2} \end{eqnarray} Equations (19) and (20) are purely theoretical in form. If we have an optical system which images the cloud onto a CCD, then we need to consider the magnification of the system as well as the pixel size. The cloud size will be measured in the number of pixels. Each pixel has a size, $l_{pixel}$, which when multiplied measures the actual size of the object produced by the optical system. To get the actual size, we then divide by the magnification, $M$, of the system. Thus, equation (20) becomes: \begin{eqnarray} \sigma_{actual}(t)= \frac{l_{pixel}}{M}\sigma(t) = \frac{l_{pixel}}{M}\sqrt{\frac{k_BT}{m}t^2 + \sigma_0^2} \end{eqnarray}
\subsection{Geometric optics of microscopes}
A typical system that is used in absorption imaging (see Figure 1) consists of two lenses that are used to focus an image onto a CCD. Such two lens system is generally classified as microscopic as it either magnifies or demagnifies the resultant image relative to the object. The magnification depends on a number of parameters of the system and is not exclusive to the lenses alone. It is therefore important to analyse such system and understand its behaviour under various conditions. \\ \\ We consider two perfect lenses placed at a distance $f_1 + f_2$ between each other. An object is placed at a distance $s=f_1$ in front of the first lens (as shown in Figure 1). We apply the Thin Lens equation to the first lens: \begin{eqnarray} \frac{1}{f_{1}}=\frac{1}{s_{1}} + \frac{1}{s_{1}^{'}} \end{eqnarray} But $s=f_1$, hence we have that $s_{1}^{'} = \infty$. $s_{1}^{'} = \infty$ implies that the beam emerging on the right hand side of the first lens is collimated.
\begin{center}
\includegraphics{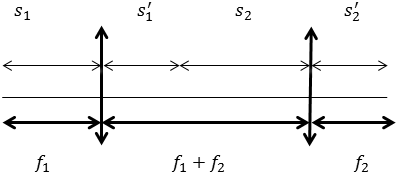}
\end{center}
Figure 1: Two lens microscopic system presented in an ideal scenario where the object and the image are at the focal lengths of both lenses. \\ \\ Now, we want to find the distance $s_2$, from Figure 1 this is just: \begin{eqnarray} s_2 = f_1 + f_2 - s_{1}^{'} \Rightarrow s_2 = f_1 + f_2 - \infty \Rightarrow s_2 = - \infty \end{eqnarray} As $f_1 + f_2 \ll \infty$. Now applying the Thin Lens equation to the second lens we get: \begin{eqnarray} \frac{1}{f_{2}}=\frac{1}{s_{2}} + \frac{1}{s_{2}^{'}} \Rightarrow \frac{1}{f_{2}}=\frac{1}{-\infty} + \frac{1}{s_{2}^{'}} \Rightarrow f_2 = s_{2}^{'} \end{eqnarray} So from this we can conclude that if the object is placed at a focal length of the first lens and the distance between the two lenses is the sum of their focal lengths, then the final image will form at a focal length of the second lens. \\ \\ Conversely, if the beam entering the system is infinity corrected, i.e. $s=\infty$, then it will emerge parallel to the optical axis as shown in figure 2.
\begin{center}
\includegraphics{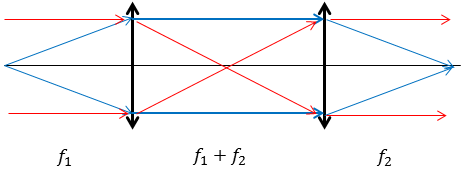}
\end{center}
Figure 2: Ray tracing plot of two lens system. Blue: the object is placed at a focal length of the first lens. Red: the beam entering the first lens is parallel to the optical axis. \\ \\ Note that if the distance between the two lenses is not, $f_1 + f_2$, then the emerging beam on the right hand side of the OS will no longer be collimated. In other words, the beam will be converging or diverging depending on whether the distance between, $f_1 + f_2$, is bigger or smaller. \\ \\ The total magnification of an $n$ lens system is just the product of magnifications of each lens: \begin{eqnarray} m_T=\prod \limits_{i=1}^n m_i = \prod \limits_{i=1}^n (-1)^n \frac{s_{i}^{'}}{s_i} \end{eqnarray} Hence, for our two lens system it is just simply: \begin{eqnarray} m_T=\frac{s_{1}^{'}}{s_1} \times \frac{s_{2}^{'}}{s_2} \Rightarrow m_T=-\frac{f_2}{f_1} \end{eqnarray} In reality being able to put the object at exactly the focal distance is not necessarily possible. As a result it is important to understand how the behaviour of the system changes when the object is not at the focal point of the first lens. Figure 3 shows such situation where the object is placed at a distance, $s_1 = f_1 + \delta$, where, $\delta \in\forall\Re$. 
\begin{center}
\includegraphics{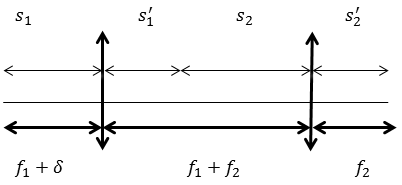}
\end{center}
Figure 3: Two lens microscopic system presented in non -- ideal scenario where the object is not at the focal point of the first lens. \\ \\ Once again we apply the Lens Maker's equation on the first lens and rearrange to get $s_{1}^{'}$: \begin{eqnarray} s_{1}^{'}=\frac{s_1 f_1}{s_1-f_1} \Rightarrow s_{1}^{'}=\frac{f_1^2}{\delta}+f_1 \end{eqnarray} Now the object distance $s_2$ from the second lens is: \begin{eqnarray} s_2 = f_1+f_2 - s_{1}^{'} \Rightarrow s_2 = f_2 - \frac{f_1^2}{\delta} \end{eqnarray} Applying the Thin Lens equation to the second lens and rearranging for $s_{2}^{'}$ and simplifying it we obtain: \begin{eqnarray} s_{2}^{'}=-\frac{\delta f_2}{f_1^2}(f_2-\frac{f_1^2}{\delta}) \end{eqnarray} The final image no longer forms at the focal point of the second lens. However, as $\delta \rightarrow 0$, then the equation (28) tends to $f_2$ as expected. Finally, using equation (24) we obtain the total magnification of the system: \begin{eqnarray} m_T = \frac{s_{1}^{'}}{s_1} \times \frac{s_{2}^{'}}{s_2} = \frac{\frac{f_1}{\delta}(f_1+\delta)}{f_1+\delta}\times\frac{-\frac{\delta f_2}{f_1^2}(f_2 - \frac{f_1^2}{\delta})}{f_2 - \frac{f_1^2}{\delta}}\end{eqnarray} The expression above reduces to: \begin{eqnarray} m_T=-\frac{f_2}{f_1} \end{eqnarray} From this we can infer that changing the distance between the first lens and the object has no effect on the total magnification of the system. \\ \\ We now proceed to consider an even more complex case. As in previous part we assumed that the object is placed at a distance that is not equal to the focal length of the first lens. Now, we will also assume that the distance between the two lenses is not $f_1 + f_2$, but $f_1 + f_2 + \alpha$, where, $\alpha$, is some deviation from the ideal distance. Figure 4 depicts the situation.
\begin{center}
\includegraphics{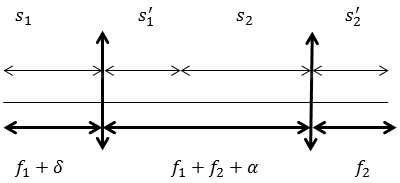}
\end{center}
Figure 4: Two lens microscopic system presented in non -- ideal scenario where the object is not at the focal point of the first lens and the two lenses are not at a distance $f_1 + f_2$. \\ \\ For the first lens, as previously, the distance $ s_{1}^{'}$ is the same, namely: \begin{eqnarray} s_{1}^{'}=\frac{f_1^2}{\delta}+f_1 \end{eqnarray} The new distance $s_2$ is then: \begin{eqnarray} s_2=f_1+f_2+\alpha - s_{1}^{'} = f_1+f_2+\alpha - \frac{f_1^2}{\delta}-f_1 \Rightarrow s_2 = f_2 + \alpha - \frac{f_1^2}{\delta} \end{eqnarray} Applying the Thin Lens equation to the second lens and rearranging to get $s_{2}^{'}$ we get: \begin{eqnarray} s_{2}^{'}=\frac{s_2f_2}{s_2-f_2}=\frac{(f_2+\alpha -\frac{f_1^2}{\delta})f_2}{\alpha-\frac{f_1^2}{\delta}} \end{eqnarray} Once again we observe that the new image no longer forms at the focal length of the second lends but instead at some distance that depends on the parameters $\alpha$ and $\delta$. Once again, in the limits as $\delta \wedge \alpha \rightarrow 0$, the object is formed at the focal length of the second lens, as expected. \\ \\ The total magnification of the system is: \begin{eqnarray} m_T=\frac{f_1f_2}{\alpha\delta -f_1^2} \end{eqnarray} In this case we observe that the magnification is not $-f_2/f_1$ as in previous case. The deviations $\alpha$ and $\delta$ couple. \\ \\ The expression above describing the effects of magnification based on variations in distance has got some important consequences. First, if the distances $\alpha$ and $\delta$ are large, the magnification is unstable for large $\alpha \delta$. Hence, if our object's position cannot be fixed and can experience some variation in $\alpha$, then it's resultant magnification would vary. This instability would have consequences in the calculations of temperature of the cloud since our model assumes that the magnification is fixed. Therefore, optimising the object to first lens and lens to lens distances is of prime importance as it minimises the $\alpha \delta$ product and reduces the magnification instability even if the position of the object varies from time to time in the TOF experiments. In reality, the lens to object distance accuracy is harder to achieve, therefore, if the lens -- lens distance carries a small deviation, $\delta$, then the coupling term in equation (35) vanishes as it becomes much smaller than $f_1^2$.
\subsection{Resolution of a two--lens system}
An important quantity in optical design process is the numerical aperture, NA. The numerical aperture is a dimensionless quantity that defines the range of angles that an optical system can accept or emit light. This is given by: \begin{eqnarray} NA = nsin\theta=nsin(arctan(\frac{D}{2f})) \end{eqnarray} Where, $n$, is the refractive index of the medium, $D$, is the diameter of the lens and, $f$, is the focal length of the lens. The diffraction limited resolution can be estimated using the well known relation: \begin{eqnarray} x_{min}=\frac{0.61\lambda}{NA} \end{eqnarray} Where, $\lambda$, is the wavelength of light used. For multiple lens systems, the resolution of the system is dictated by the first lens (lens closest to the object that is being imaged). The closer the lens is to the object, the higher is the light gathering power and so the information loss is reduced. In addition, each lens is not perfect and therefore has some underlying impulse response or point spread function, (PSF), associated with it. The final image for an, $n$, lens system in Fourier plane is given by the convolution between the object and the corresponding point spread functions of each lens: \begin{eqnarray}I(x,y)=O(x,y)\cdot PSF_1(x,y)\cdot PSF_2(x,y)\cdot\ldots \cdot PSF_n(x,y) \end{eqnarray} In equation (35) the order of multiplication matters i.e. it is not commutative if the object and the point spread functions are expressed in finite matrix arrays.
\subsection{RIC, OPD and MTF curve characteristics}
There are a number of indicators in computational optics which give information on image quality and the overall performance of optical system. Understanding these is therefore key in being able to design an optical system free of defects, low aberrations and good resolution. \\ \\ Ray intercept curves, or ray trace curves are prime indicators of the aberrations that are present in the optical design. They can qualitatively tell the type of aberrations that exists in the optical design and also give a quantitative representation of how big they are [4]. Figure 5 below shows an example of a RIC for a microscope lens system.
\begin{center}
\includegraphics{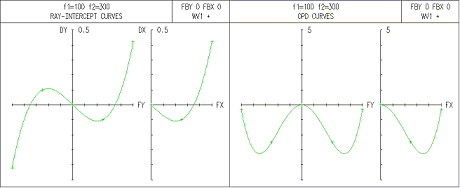}
\end{center}
Figure 5: RIC and OPD curves for an $f_1=100mm, f_2=300mm$, microscope system using OSLO. \\ \\ Here the aberrations are $3^{rd}$ order spherical aberrations characterised by a cubic polynomial. The image was placed at an optimum focus that minimises the residual blur diameter. Spherical aberrations are characterised by a variation in focus position with aperture. \\ \\ Optical path difference, (OPD), is another important quantity in optical design. It quantitatively describes the difference between a perfect wavefront and a real wavefront [4]. Generally, if the OPD is $\frac{1}{4}\lambda$, then the system is diffraction limited. Here, in the figure above, we can see that the maximum OPD is more than $3\lambda$, which exceeds the diffraction limit. The distortion of the wavefront varies spatially across the optical system. Optical path differences and the distortion of the wavefront depend on the quality of the lens and particularly its surface smoothness. \\ \\ The final and most comprehensive indicator in optical performance that will now be discussed is the modulation transfer function, MTF. Figure 6 shows a computed MTF curve for $f_1=100mm, f_2=300m$, microscope system using OSLO. 
\begin{center}
\includegraphics{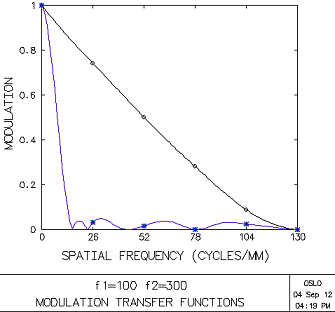}
\end{center}
Figure 6: computed MTF curve for $f_1=100mm, f_2=300mm$, microscope system using OSLO. Here the monotonically decaying straight lines represents how the MTF would behave in ideal situation and the blue curve represents what the performance of the system actually is. \\ \\ The modulation transfer function is estimated by comparing the contrast of the image with that of the object. Namely, if we have an object and measure its minimum and maximum intensity, we can obtain its modulation [4]: \begin{eqnarray}Modulation=\frac{I_{max}-I_{min}}{I_{max}+I_{min}} \end{eqnarray} We can then place the object to be imaged by our optical system and perform the same procedure of measuring the different intensities and estimate the modulation. Finally, to calculate the modulation transfer function we take the ratio of the modulations in image and object, namely: \begin{eqnarray} MTF=\frac{MOD_{img}}{MOD_{obj}} \end{eqnarray} By performing such procedure for different size objects, we can compute the entire spectrum of modulation transfer function based on the object's size. The MTF is affected by diffraction, and geometrical aberrations. In addition, the MTF curve characteristics can vary greatly across different spatial frequencies, i.e. a fairly large object can have a lower modulation than a smaller object of higher spatial frequency. \\ \\ The cutoff frequency in the MTF is the highest possible frequency at which the modulation goes to zero -- the contrast of the system is effectively reduced to zero and the different features are of the same intensity, $I_{min}=I_{max}$. This allows one to determine the resolution of the system just by reading off the value for the cutoff frequency from the MTF, the resolution is given by the inverse of cutoff frequency. 
\section{Optical design and development}
\subsection{Initial stages: establishing constraints and optical performance requirements}
Before the actual design of the system takes place, there must exist a period of development whereby the designer establishes what it wants to achieve with the optical system and what sort of constraints limit the optical design. Based on these limitations and demands one can then start the process of initial development where mathematics and computer simulations take place to test the performance of the system. Finally, given that the system passed all the initial development tests it is then sensible to realise the design in a real experimental environment and test it. \\ \\ The optical system design and optimisation procedure can be summarised in the figure 7. We begin to describe each step in the flow chart in more detail [4]. \begin{enumerate}
\item The first step in the design process should involve acquiring all of the specifications. These may include optical specifications such as the focal length, lens diameter, environmental requirements such as temperature and space constraints, performance goal -- what resolution or magnification one wants to achieve.
\item The starting point would be a basic configuration that can meet the specifications of the optical system to some degree. 
\item Establishing variables and constraints. The system variables generally include things like radii, thickness, airspaces, glass characteristics, wavelength of light used. The lens dimensions are usually restricted by the manufacturer's standards and cannot be varied unless a custom built design is requested. It is useful to assess the importance and impact of each variable in the optical design process and to which extent it can be varied in order to achieve the required system specification.
\item The simple prototype of the system can be programmed using optics software (e.g. OSLO or Zeemax).
\item The optimisation process of the system can have a number of requirements. For example, the final focus of the image could be optimised such that it minimises the optical path differences in the wave front or minimising the spherical aberrations by means of introducing lenses with glasses of different refractive indices. Having a detailed description of points 1 and 3 can greatly improve the optimisation procedure.
\item A number of optical indicators such as the Modulation transfer function (MTF), the Ray intercept curves (RIC), the optical path difference curves (OPD), rms sport radius and others can provide with a very comprehensive overview about the performance of the optical system. It is therefore crucial that their properties are well understood. 
\item If the performance of the system is not satisfactory, points 3, 4, and 5 can be repeated until the desired performance is achieved. 
\item In reality, it is highly improbable that the system will be built with the level of precision and alignment as specified in the simulations. In addition, the optical materials and physical conditions may not be perfect and homogenous throughout. It is therefore important to investigate the extent to which the performance of one’s optical system changes with different errors. If the tolerances appear to be significant, action can be taken early in the development process to minimise the risk in propagation of such design imperfections in the final stages of assembly.
\item Provided that the system has passed all the initial tests it can now be assembled and investigated further in real life.
\end{enumerate}
\begin{center}
\includegraphics{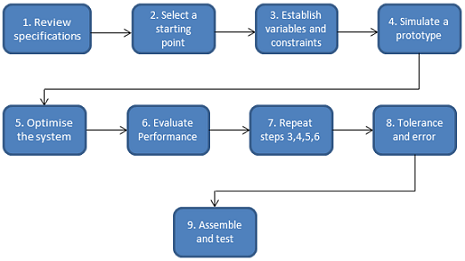}
\end{center}
\begin{center}
Figure 7: Optical system design and optimisation flowchart. 
\end{center}
\subsection{System constraints in our optical design}
We now proceed to outline some of our optical design constraints and requirements. \\ \\ 
\textbf{Physical Constraint $\sharp 1$}. The chamber size put a constraint on the focal length of the first lens such that $f_1 > 137.7mm$, this meant that we had to use a minimum of $f_1 \geq 150mm$ (minimum focal length sold by the manufacturer based on this constraint -- Thorlabs). Of course, a multiple lens system could be designed to match more closely to the distance 137.7mm, however, this will certainly add several complications in the design process. One of such complications would be the care and precision required to assemble multiple lenses systems -- this in effect would add more degrees of freedom and without the machinery human error is an inevitable end. In addition, custom lens designs from the manufacturers is very costly and is generally not considered unless the system would have a huge improvement on the performance. In short, the best optical system is one which satisfies Occam's razor. \\ \\
\textbf{Physical constraint $\sharp 2$}. The diameter of the imaging beam has also put a constraint on the size of our lenses that we had to choose. The imaging beam diameter was chosen to be $d=20mm$. Hence, to reduce the clipping of the light and thus minimise the information loss from the image, the diameter of the lenses was chosen to be $d_{lenses} = 50.4mm$, this size was limited by the manufacturer. \\ \\ 
\textbf{Optical constraint $\sharp 1$}. The magnification of the system was chosen to be around $M > 3$. This put a constraint on the choice of our second lens. A lens of $f_2 = 500mm$ was chosen to produce a theoretical magnification of $M = 3.33$. The choice of the focal length of the second lens once again was limited by the manufacturer’s standards (Thorlabs). \\ \\ 
\textbf{Physical constraint $\sharp 3$}. The above optical constraint has further implicated the space constraint. The significant increase in length of the optical system meant that it could no longer be built in a linear assembly fashion. Whilst the testing could be easily carried out in a linear setup, the assembly and implementation of the system in the presence of the UHV chamber could no longer be supported by a linear setup. This is fortunately not a significant issue as it can be easily overcome by implementing mirror joints which can effectively \textquotedblleft fold\textquotedblright the optical system into a compact form without suffering degradation in the overall performance. \\ \\ There is another alternative which can be used to reduce the effective size of our optical system. In the mathematical analysis (Section 2.3) we assumed that we are imaging a mathematical point placed at the focal distance from the first lens. This implied that the emerging beam from the second lens was collimated. As a result, the second lens could be placed very close to the first one and still form an image at a focal point of the second lens. In reality, this can be achieved as far as the paraxial approximation is satisfied. However, as discussed in section 2.3, it adds further complications such as the magnification instability and the fact that the laser beam is no longer collimated. \\ \\
\textbf{Optical constraint $\sharp 2$}. The operating imaging beam wavelength was set to $\lambda = 780.032nm$. This put a constraint on choosing the wavelength operating range for our lenses. Thorlabs provides $\lambda = 650 - 1050nm$ operating range. \\ \\
\textbf{Optical constraint $\sharp 3$}. Since our imaging system would be composed of two lenses it was important to establish what type of lenses to use. In the initial development, plano convex lenses were used. However, it was found that they carried large spherical aberrations and optical path differences. The next best option was found to be using achromat doublets. These lenses are made out of two different types of glass -- flint and crown, as a result, they significantly reduced the spherical aberrations and the optical path differences.\\ \\ \textbf{Optical Constraint $\sharp 4$}. Using the information of the known optical constraints and the theory on the cloud expansion rates at a given temperature (Refer to 2.2 Derivation of the temperature of the cloud) we can estimate the constraints which our imaging camera must satisfy. Taking the standard mass for Rubidium to be $m_{Rb}=85.4678u$, where, $u=1.660538921(73)\times 10^{-27}kg$, is the unified atomic mass, temperature, $T\simeq 30\mu K$, and allowing the time of flight expansion to be $30ms$, we estimate (eqn 19) the size of the cloud, $\sigma=1.61mm$. In addition, the cloud will fall under gravity for $30ms$ covering a distance of $s=\frac{1}{2}gt^2=4.41mm$. Using this information we can estimate the extremum of the cloud edge, this is given by the sum of the cloud size and the distance moved: $d=6.02mm$. However, our system has a magnification that is $M=3.33$, and so the resultant image projected onto the CCD will be larger by the same factor. Hence, our active surface area should be no less than $20mm\times 20mm$. In addition, the pixel size of our camera is constrained by the resolution of our imaging system. Therefore, to utilise the achieved resolution, one should attempt to choose a camera with a pixel size that is smaller than the given resolution, but at the same time not too small so that quantum efficiency is not affected too greatly (quantum efficiency of photon detection goes down as the pixel size decreases).
\subsection{Development and Simulations}
The initial system development was carried out using OSLO software (Optics Software for Layout and Optimization). The lens material specifications and dimensions can be readily entered into OSLO and optimised using inbuilt algorithms. \\ \\ The simulation procedure involved the following steps: 
\begin{enumerate}
\item Build the first lens. Optimise the lens distance such that the distance between the object and the lens produces a collimated beam on the right hand side. The optimisation procedure would be carried out using aberration operands, more specifically, using, PU, or paraxial axial ray optimisation operand. We have included the chamber window in the optical simulation to account for any effects due to the extra optical components. The beam radius was set to, $r=10mm$, wavelength, $\lambda=780.032nm$. The achromat doublet was placed with the flatter end (large radius of curvature) to face the object. This reduces the level of spherical aberrations and distortion of the wavefront. 
\begin{center}
\includegraphics{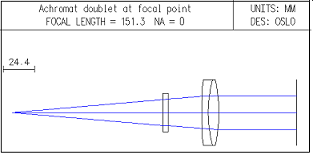}
\end{center}
Figure 8. Achromat doublet with the mathematical point at the focal point. The optimisation was carried out using paraxial axial ray optimisation.
\item A second lens is placed at an arbitrary distance from the first lens. The object is now placed at an infinity (the rays are collimated). Once again, we use Paraxial--axial ray optimisation algorithm to optimise the distance $f_1 + f_2$ so that the emerging beam is collimated on the right hand side of the optical system. The second lens is placed such that it faces the first lens with a surface of smaller radius of curvature. Figure 9 displays the setup in OSLO.
\begin{center}
\includegraphics{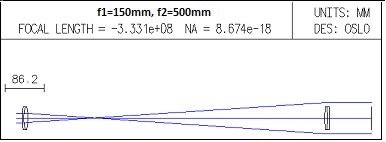}
\end{center}
Figure 9: A pair of achromat doublets with image placed at infinity. Here $f_1=150mm$, and $f_2=500mm$. \\ \\ From this simulation we can observe that the emerging beam from the first lens has a diameter that is larger than the second lens. This implies that there will be some clipping of light involved. Clipping of light results in the loss of information about the image and a lower signal to noise ratio. To remedy this there are a number of options. First, we could increase the diameter of the second lens so that all of the light is collected. However, in our case we cannot do that since the manufacturer does not offer lenses with a bigger diameter. Secondly, we could sacrifice the magnification stability by reducing the distance between the two lenses. However, in our case that could seriously affect the calculations of the atom number and temperature of the cloud.
\item Finally, the optimised distances are combined in a system were the object is placed at a focus of the first lens and the distances between the two lenses are placed at the optimised distance $f_1 +f_2$. The final image is focused using an Autofocus optimisation option for minimum optical path difference for monochromatic light. Figure 10 shows the final optimised setup and the table below summarises the optical component distances.
\begin{center}
\includegraphics{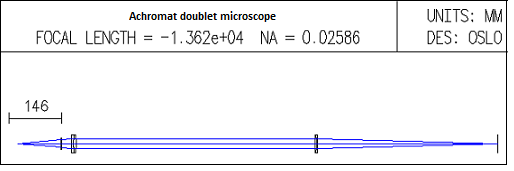}
\end{center}
Figure 10. Optimised achromat doublet microscopic system for ultracold atom imaging.
\end{enumerate}
\begin{table}[ht]
\caption{Simulation results} 

\centering 

\begin{tabular}{c c c c c} \\ 
\hline\hline 

$s_1=f_1$, (mm) & $f_1+f_2$, (mm) & $s^{'}_2=f_2$, (mm) \\ [1ex] 
\hline 

146.5 & 654.8 & 494.3 & \\ 
[1ex] 
\hline \hline 
\end{tabular}
\end{table}
We now begin to analyse the performance of our optical system by looking at the parameters such as the RIC, OPD, MTF. Figure 11 displays the RIC and OPD curves.\\ \\ From the OPD curve we can see that the wavefront distortion is nearly $\simeq\frac{1}{4}\lambda$ -- which is close to the Rayleigh limit. In addition, the spherical aberrations are $3^{rd}$ order spherical aberrations. If we were to compare this level of aberrations with an identical achromat doublet system, but with their faces reversed (i.e. smaller radius of curvature facing the object), we would find that the aberrations are 6 times smaller with an OPD maximum being 4 times smaller in the first system compared to the second one.
\begin{center}
\includegraphics{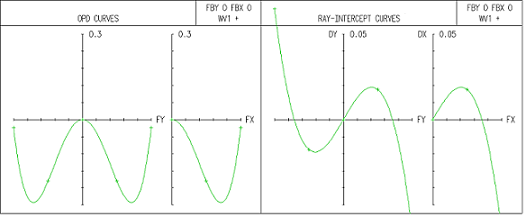}
\end{center}
Figure 11: Optical Path difference and Ray intercept curves for the achromat doublet system, $f_1=150mm, f_2=500mm$. \\ \\ The modulation transfer function of our system has the following behaviour depicted in figure 12. The MTF curve of our system follows closely the behaviour of the ideal system which indicates small geometric aberrations. This is confirmed by the OPD and RIC curves. \\ \\ The resolution of the system can be estimated by reading off the cutoff frequency from the MTF curve, namely: \begin{eqnarray} x_{min}=\frac{1}{\nu_{cutoff}} \end{eqnarray} Taking the cutoff frequency to be approximately equal to $\nu_{cutoff}\simeq 70$ $cycles$ $mm^{-1}$. With this value, we can estimate the theoretical resolution over one cycle which is defined as a distance between the brightest point and the darkest point.
\begin{center}
\includegraphics{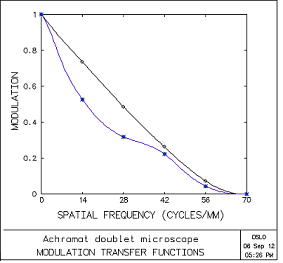}
\end{center} 
Figure 12: MTF curve for the achromat doublet system, $f_1=150mm, f_2=500mm$. Here the black curve depicts the MTF of an ideal lens system and the blue curve represents how our lens system should behave in reality. \\ \\ This gives us a theoretical estimate for the resolution of the system: \begin{eqnarray} x_{min}=\frac{1}{\nu_{cutoff}}=\frac{1}{70}= 14.3\mu m \end{eqnarray} If we were to base our resolution estimate on equation (37), then we would find that it yields: \begin{eqnarray} x_{min}=1.22\frac{\lambda f}{D}= 1.22 \frac{780.032nm\cdot 150mm}{30mm}=4.76\mu m \end{eqnarray} In our case, the best resolution that can be achieved is limited by the chamber window (chamber window clear aperture diameter $D=30mm$). \\ \\ We can observe that the two estimates show a substantial difference. This can be explained, at least in some part, by the fact that the spatial resolution estimate using equation (43) does not take into account any geometrical aberrations that may be present in the system. \\ \\ The magnification of the system given by OSLO was $M=3.31$.
\subsection{Tolerances and Errors}
Whilst it is possible to achieve a perfectly optimised system in the simulation, in real experimental conditions this is not the case. The optimum distances between the optical components and object/image may not necessarily be achieved. As a result, it is important to understand how well our optical design can tolerate such imperfections or errors and how much it can affect the image quality. \\ \\ First we proceed to investigate the tolerancing effects of the object being not at the optimum focus of the first lens. Increasing the object--lens distance we observed astigmatism in the ray intercept curves. The wavefront distortion increased significantly with the wavefronts having largest optical path differences near the edges. The performance described by the MTF curve also showed negative effects, though the resolution of the system remained unchanged. Figure 14 summarises the tolerancing effects. 
\begin{center}
\includegraphics{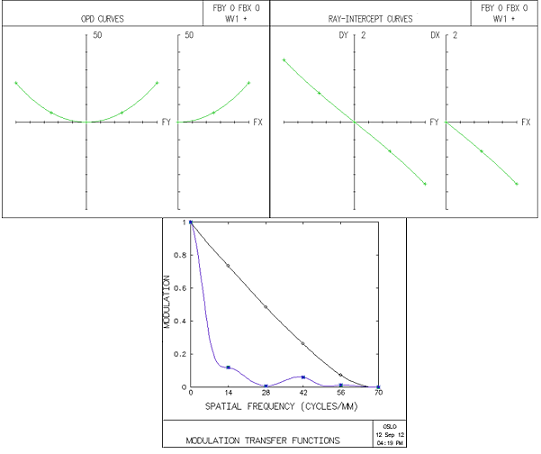}
\end{center}
Figure 13: System performance when we place the object at $s=f_1+\delta=150+5mm$. \\ \\ Next, we move the object closer to the first lens. This time, we observe astigmatism and defocus in the ray trace curve. The wavefront distortion follows the same profile but is reflected in the x-axis. The MTF curve once again shows a decrease in performance, though no noticable change in resolution was observed. Figure 14 shows the results. \\ \\ In both cases the system displays similar characteristics, however, the MTF curves differ due to different profiles of geometric aberrations. We now proceed to vary the distance between the two lenses as well as the distance between the object and the lens. 
\begin{center}
\includegraphics{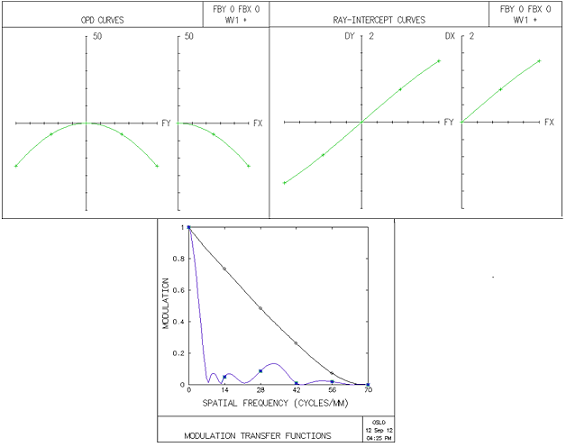}
\end{center}
Figure 14: System performance when we place the object at $s=f_1+\delta=150-5mm$. \\ \\ The tolerancing of the system when the lens--object and lens--lens distances were varied simultaneously did not show any noticeable changes in geometric aberrations, i.e. the geometric aberrations and the performance of the system remained identical to the case were only object--lens distance was varied (in each case we varied the distances by up to $\pm 5mm$). However, a change in magnification was observed. The magnification varied by few parts in hundreds of decimal point. This is expected from the analysis carried out in section 2.3. \\ \\ Now we look at how the system performance varies when the lens--camera distances are not at their optimum points. Once again these are varied by $\pm 5mm$. Figure 15 depicts the results. \\ \\ We can observe that the aberrations and wavefront distortion is much smaller relative to the errors produced by the first lens. The aberrations in the system remain $3^{rd}$ order spherical aberrations. The optical path differences have increased, though are by a factor of 10 smaller than the ones produced by the first lens. The MTF curve seems to have poor performance in the intermediate regime, however, for small features it performs as an ideal system. 
\begin{center}
\includegraphics{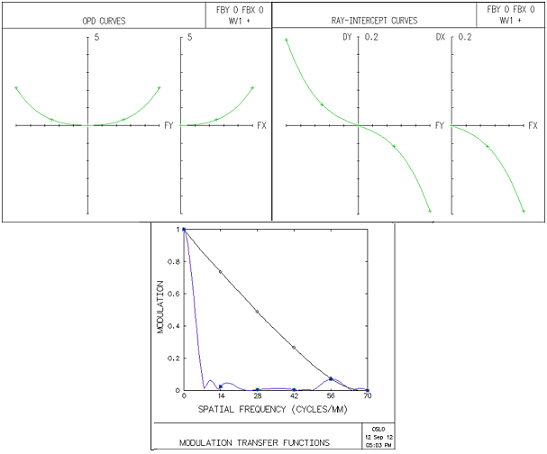}
\end{center}
Figure 15: System performance when we place the camera at $f_2+\delta=500+5mm$ \\ \\ Qualitatively, when the distance between the lens and the camera is decreased by the same amount, the aberrations remain spherical and the OPD maximum is reduced by a factor of 2. The MTF curve performed poorly across different spatial frequencies. \\ \\ When both object--lens and lens--camera distances were varied the aberrations and optical path differences including the MTF performance are completely dominated by the effects produced by the first lens. 
\section{Building and Testing}
\subsection{Lateral Imaging}
The first part of imaging system testing was in the lateral imaging configuration. This is where the collimated laser beam incident on the object is completely parallel to the optical axis. We have used single mode optical fibre fed into a collimator as an imaging beam. \\ \\ The setup configuration was realised on an optical bench with optical components being mounted on rails and linear translation stages. As an imaging target, we have used a 1951 USAF target. For imaging, we have selected a Guppy Pro F--125 camera with $1292\times 964$ resolution and $3.75\mu m$ pixel size. Figure 16 depicts the experimental setup. 
\begin{center}
\includegraphics{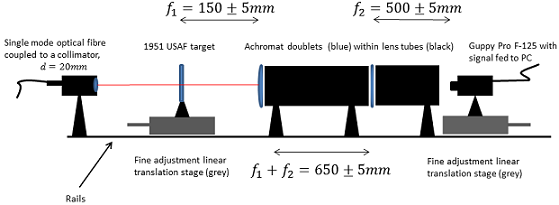}
\end{center}
Figure 16: Experimental setup for testing the lateral imaging system (not to scale). We used 7T175-50 - Aluminium Translation Stages with 50 mm travel range.\\ \\ With this experimental arrangement we have obtained the following image shown in figure 17. The image has some interference effects caused by coherent and monochromatic laser light. Some of the interference is produced by small debri on lenses and the target. \\ \\ To determine the magnification of the system, we selected a square feature (marked blue on figure 17) which was the $2^{nd}$ element in group 4. The square covered 2.5 cycles, and it had a spatial frequency of 17.9 cycles $mm^{-1}$ (as read from the standard 1951 USAF size chart). From this it was possible to calculate the size of the square over 2.5 cycles which gave 0.140mm. The 0.140mm is the size of the square if the magnification of the system is $M=1$. Using ImageJ software the length of the square was measured to be $126\pm 1$ pixels in x and y direction (to nearest pixel). 
\begin{center}
\includegraphics{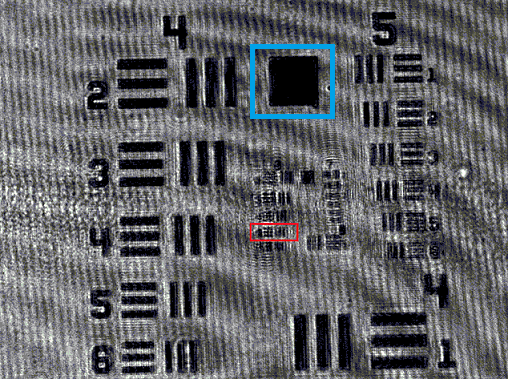}
\end{center}
Figure 17: 1951 USAF target using the lateral imaging configuration. The red square indicates the smallest feature that could be resolved on the target whilst the blue feature was used to estimate the magnification of the system. \\ \\ Knowing the pixel size, ($3.75\mu m$), the size of the magnified image could be calculated. It is simply the product of pixel size and number of pixels measured namely: $length=126 \pm 1 pixels \times 3.75\mu m=0.473\pm 0.004mm$. The error has been obtained by taking repeated measurements of length and estimating the standard error on the mean. This error calculation is not extensive as it does not take into account any systematic, rounding or alignment errors. Now, to estimate the magnification of the system, we divide the experimentally estimated size by the theoretical estimate of magnification 1: \begin{eqnarray}M=\frac{length_{exp}}{length_{theo}}=\frac{0.473mm\pm 0.004}{0.140mm}=3.38 \pm 0.03 \end{eqnarray} Now the smallest feature that could be resolved in the x and y direction (marked red) was element 5 in group 6. From the standard 1951 USAF target this had a spatial frequency of 102 cycles $mm^{-1}$. With this value, the resolution of the system is: \begin{eqnarray} x_{min}=\frac{1}{\nu_{cutoff}}=\frac{1}{102cycles mm^{-1}}=9.80\mu m \end{eqnarray} The standard convention in estimating the resolution using a 1951 USAF target requires the features in horizontal and vertical direction to be resolved and the resolution is defined over one full cycle which is the distance between the bright and dark fringe. \\ \\ The experimental estimates of resolution and magnification are relatively similair to the estimates obtained in the theoretical calculations and simulations, though there are some differences. 
\subsection{Grazing Incidence Imaging}
In grazing incidence imaging, the collimated imaging beam is reflected off an atom chip (in our case mirror). If an object is close enough to the mirror, a virtual image is produced and the final image consists of two objects. The distance between the two objects allows one to determine the chip--object distance. \\ \\ To simulate an atom chip we have used a 50.8 mm protected gold mirror from Thorlabs. It was mounted parallel on a separate rail. For fine adjustment, the mirror was attached to a 6mm range, 7T67-6 -- Stainless Steel Translation Stage from Standa. \\ \\ It was not practical to implement the 1951 USAF target since the smaller features were far away from the mirror. Instead, the target was replaced by a single hair of $0.05\pm 0.01mm$ thickness. The imaging beam was tilted using fine adjustment screws. A protractor was used to measure the angle of inclination. Figure 18 depicts the experimental setup. 
\begin{center}
\includegraphics{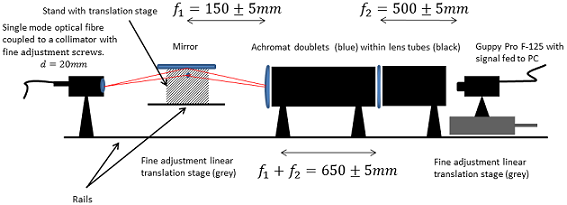}
\end{center}
Figure 18: Experimental setup to test grazing incidence imaging (not to scale). \\ \\ In grazing incidence imaging, we tilted the beam by $1^{\circ}-2^{\circ}$ degrees. The resultant images obtained are depicted in figure 19.
\begin{center}
\includegraphics{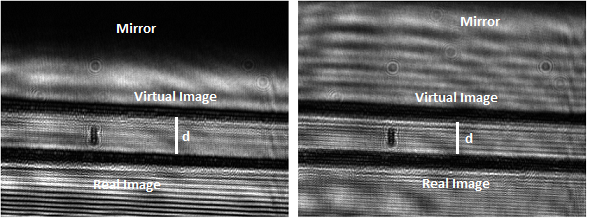}
\end{center}
Figure 19: Grazing incidence imaging. Left image, $\theta = 1^{\circ} \pm 0.5^{\circ}$. Right image, $\theta = 2^{\circ} \pm 0.5^{\circ}$. The interference fringes in the images are produced due to diffraction of light around the edges of the mirror. \\ \\ It must be noted that the mirror blocks some of the light incident on the object and so the signal to noise ratio increases. \\ \\ To determine which one of the images is virtual and which one is real, the object was moved and the corresponding motion of the two images observed. The real image moves in the same direction as the object moved, whereas the virtual image moves in the opposite direction. \\ \\ The distance, $d$, between the virtual and real image can be measured and used to calculate the distance of the object away from the chip. In terms of geometrical optics, some of the imaging beam is first incident on the chip and then reflected back onto the image which produces a shadow -- this is the real image. In the second case, the imaging beam is first incident on the object which then casts a shadow on the chip which is then reflected off it generating a virtual image. 
\begin{center}
\includegraphics{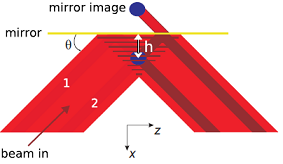}

Figure 20: Ray propagation in grazing incidence imaging [1]. 
\end{center}
Using trigonometry, the distance, $d$, can be estimated using [1]: \begin{eqnarray} d=2hcos\theta \end{eqnarray} Hence, to find the chip--object distance, $h$, we can just rearrange the equation above. It must be noted, that if the object gets too close to the chip, then the virtual and real images will merge together. This of course depends on the size of the object being imaged as well as the resolution of the system. 
\section{Conclusion}
In this report I have discussed the technical details on how to develop and build an absorption imaging system for ultracold atoms. The optical design, whilst relatively simple, has numerous capabilities in probing and extracting information from cold atoms ranging from the temperature calculations to the distances from the chip. However, this simplicity appears to be superficial and various technical intricacies begin to emerge when system constraints are taken into account. \\ \\ It is evident that, in order to improve the performance of the optical design one should attempt to ease the optical and physical constraints as these primarily limit the choice of the optical components and ultimately determine the resolution of the system. 
\section*{Acknowledgments}
The author would like to thank Fedja Oru\u{c}evic, Stefano Bettelli, Bo Lu, Samanta Piano, Peter Kr\"{u}ger and Thomas Fernholz for helpful discussions. This work has been supported by ESPRC vacation bursary scheme. 
\section*{References}
$[1]$ David A. Smith, et.al., \textquotedblleft Absorption imaging of ultracold atoms on atom chips", Optics Express, Vol. 19, Issue 9, pp. 8471-8485 (2011).\\ $[2]$ Daniel A. Steck, \textquotedblleft Rubidium 87 D line data", (2001). \\ $[3]$ H. J. Lewandowski, et.al., \textquotedblleft Simplified System for Creating a Bose-Einstein Condensate". \\ $[4]$ Robert E. Fischer, et.al., Optical System Design, Second Edition, McGraw--Hill, (2008).
\end{document}